\documentclass{josis}
\usepackage{hyperref}
\usepackage[hyphenbreaks]{breakurl}
\usepackage{booktabs}
\usepackage{stmaryrd}
\usepackage[T1]{fontenc}
\usepackage{cite}

\usepackage{algorithm}
\usepackage{algpseudocode}

\usepackage[table]{xcolor}
\usepackage{amssymb,amsmath}

\usepackage{mathtools}
\usepackage{xfrac}
\usepackage{siunitx}
\usepackage{setspace}
\usepackage{soul}

\DeclareSIUnit{\belmilliwatt}{Bm}
\DeclareSIUnit{\dBm}{\deci\belmilliwatt}

\newcommand{\defn}[1]{\emph{#1}}    

\newcommand{\RR}{\mathbb R}
\newcommand{\Prob}{\mathbb P}     

\newcommand{\Vor}{\text{Vor}}
\newcommand{\Ann}{\text{Ann}}
\newcommand{\strength}{\text{strength}}
\newcommand{\composite}{\text{composite}}
\newcommand{\uniform}{\text{uniform}}
\newcommand{\landuse}{\text{land use}}
\newcommand{\network}{\text{network}}
\newcommand{\dist}{\text{dist}}
\newcommand{\azi}{\text{azi}}
\newcommand{\elev}{\text{elev}}
\newcommand{\steep}{\text{steep}}
\newcommand{\midp}{\text{mid}}

\graphicspath{{figures/}}

\josisdetails{%
   number=N, year=YYYY, firstpage=xx, lastpage=yy, 
  doi={10.5311/JOSIS.YYYY.II.NNN},
   received={}, 
   returned={},
   revised={},
   accepted={}, }

\makeatletter
\let\UrlSpecialsOld\UrlSpecials
\def\UrlSpecials{\UrlSpecialsOld\do\/{\Url@slash}\do\_{\Url@underscore}}%
\def\Url@slash{\@ifnextchar/{\kern-.11em\mathchar47\kern-.2em}%
    {\kern-.0em\mathchar47\kern-.08em\penalty\UrlBigBreakPenalty}}
\def\Url@underscore{\nfss@text{\leavevmode \kern.06em\vbox{\hrule\@width.3em}}}
\makeatother

\hypersetup{
colorlinks=true,
linkcolor=black,
citecolor=black,
urlcolor=black
}

\graphicspath{{figures/}} 

\runningauthor{\begin{minipage}{.9\textwidth}\centering Tennekes, Gootzen\end{minipage}}
\runningtitle{Location estimation}

\begin{document}

\title{A Bayesian approach to location estimation of mobile devices from mobile network operator data}

\author{Martijn Tennekes}\affil{Department of Methodology, Statistics Netherlands, The Netherlands}
\author{Yvonne A.P.M. Gootzen}\affil{Department of Methodology, Statistics Netherlands, The Netherlands}

\maketitle


\begin{abstract}



Mobile network operator (MNO) data are a rich data source for official statistics, such as present population, mobility, migration, and tourism. Estimating the geographic location of mobile devices is an essential step for statistical inference. Most studies use the Voronoi tessellation for this, which is based on the assumption that mobile devices are always connected to the nearest radio cell. 

This paper uses a modular Bayesian approach, allowing for different modules of prior knowledge about where devices are expected to be, and different modules for the likelihood of connection given a geographic location. We discuss and compare the use of several prior modules, including one that is based on land use. We show that the Voronoi tessellation can be used as a likelihood module. Alternatively, we propose a signal strength model using radio cell properties such as antenna height, propagation direction, and power. Using Bayes' rule, we derive a posterior probability distribution that is an estimate for the geographic location, which can be used for further statistical inference. 


We describe the method and provide illustrations of a fictional example that resembles a real-world situation. The method has been implemented in the \texttt{R} packages \texttt{mobloc} and \texttt{mobvis}, which are briefly described.
\end{abstract}

\section{Introduction}

Mobile network operator (MNO) data have shown to be a rich potential source for official statistics, in particular on present population~\cite{deville14,ahas15,meersman16,kondor17,xu18,salgado2020b}, mobility~\cite{kung13,iqbal14,alexander15,diao15,pucci15,widhalm15,jiang16,zagatti18,dujardin20}, migration~\cite{lu16,Wilson16,lai19}, and tourism~\cite{deville14}. Such statistics can be used to support a wide range of policy issues, for instance regarding mobility restrictions in order to delay and contain the COVID-19 pandemic~\cite{grantz20,oliver20,badr20}.

Two types of MNO data are needed for statistical inference: \textit{cell plan} data, which contains information about the mobile communication network infrastructure, and \textit{event} data, which are data from transaction events between the network and mobile devices.

A mobile communication network is also called a \textit{cellular network}, where each \textit{cell} enables mobile communication for a specific land area. The cell plan specifies the geographical location of the cells and their physical properties. Two types of cells can be distinguished~\cite{panwar16}: a \textit{macro cell} that is placed in a cell tower or on top of a roof, which has a theoretical range of about 30 kilometers, and a \textit{small cell} that is used to enable mobile communication inside buildings and in dense urban areas, and has a theoretical range of two kilometers.\footnote{A small cell is a general name for \textit{micro}, \textit{pico}, and \textit{femto cells}, which have theoretical ranges of respectively 2000, 200, and 20 meters.} An important cell property is the propagation direction. A cell can be directional or omnidirectional. In practice, most real-world deployed macro cells are directional, often covering an angle of about 120 degrees whereas small cells are usually omnidirectional.

There are generally two types of event data, depending on which transaction events are contained. Event data that are used to calculate the costs in order to bill customers are called \textit{Call Detail Records} (CDR), which contain events related to active mobile phone use, such as initiating a call
\footnote{CDR data contain records about calls (initiating and receiving), SMSes (sending and receiving), and mobile data usage. Note that in several studies the term CDR is used for data that only contains call and SMS events, and alternative terms \textit{Data Detail Records} (DDR) or \textit{Event Data Records} (EDR) are used for data that also include mobile data usage events.}. Event data that also contains passive events such as location updates, are called \textit{signalling data}. These data are used by the MNO for network analysis and optimization. Signalling data are usually much richer than CDR data and are therefore recommended for statistical inference.

The geographic location of devices is important for statistical inference on MNO data. Advanced location estimation techniques are used to facilitate mobile communication, but these require radio measurement data, such as the Received Signal Strength (RSS), the Signal to Interference plus Noise Ratio (SINR), and Timing Advance~\cite{khalaf06, khalaf06b, zang10, calabrese14}. However, these data are usually not stored in event data. The only location related variable that is always available in event data is the identification number of the serving cell.

The majority of studies on mobile network data use Voronoi tessellation~\cite{deville14} to distribute the geographic location of logged events. The geographic area is divided into Voronoi regions such that each Voronoi region corresponds to the geographic location of a cell tower and each point in that region is closer to that cell tower than to any other cell tower. 

There are a few downsides to using Voronoi tessellation to estimate the geographic location of devices. First of all, it assumes that all cells are omnidirectional. As described above, large range cells are often directional. The second downside of Voronoi tessellation is that it does not take other cell properties into account, such as power, height, and tilt. Third, the coverage areas of cells overlap in reality, especially in urban areas. This is because of load balancing; if a cell has reached full capacity, neighbouring cells that also have coverage are able to take over communication with mobile devices. This means that a mobile phone is not always connected to the nearest cell nor to the cell with best signal. In urban areas, a mobile phone switches frequently between cells\footnote{There are several smart phone apps that show where the connected cell is located, e.g. Network Cell Info~\cite{Wil18a}.}. The fourth and last downside of using Voronoi tessellation we would like to mention is that it does not take into account where people are expected to be.

A couple of variations of the Voronoi algorithm have been proposed to overcome some of these limitations~\mbox{\cite{graells16, meersman16, ricciato16}}. One improvement is to shift the locations of the Voronoi points from the cell tower locations towards the direction of propagation. Alternatively, when the \defn{coverage area} is known for each cell, i.e. the area which is served by the cell, the location of the centroids of these coverage areas can be used as Voronoi points. Another improvement is to create a Voronoi tessellation for cells with a large range, and subsequently assign each small cell to the Voronoi region they are located in. The Voronoi method can be extended with auxiliary data sources, such as land use, to improve the geographic location of devices~\cite{jarv17}.


An alternative to Voronoi tessellation is a Bayesian approach. Previously proposed Bayesian approaches assume availability of SINR data or Timing Advance data~\cite{zang10,ogulenko20}. We propose a modular Bayesian framework to estimate the geographic location of devices, which consists of two main modules, namely the \defn{location prior module} and the \defn{connection module}. The former employs a priori information about where devices are expected to be to produce a \defn{location prior}, and the latter uses network information to estimate the probability of connection to cells to produce a \defn{connection likelihood}. The model is generic in the sense that various data sources and methods can be used for both of these modules. Timing Advance data can taken into account in our modular approach, but it is not required.


Other Bayesian approaches often assume a uniform distribution for the location prior~\cite{khalaf06b, zang10}. While a uniform prior generally speaking represents an "objective" prior, we argue this is not true for location estimation of mobile phones. A main contribution of this paper is the exploration of the effects of various prior distributions. The most important one is the usage of land use data. This is arguably the most straightforward option, since more devices are expected to be in certain land use categories, such as urban areas, than in other land use categories, for instance grasslands. However, any data source that contains information about where devices are expected to be can be used.

The connection likelihood describes the estimated probability that a device is connected to a certain cell given its actual location, taking potential overlap between cell coverage areas into account. For this component, we propose a signal strength model which models the propagation per cell using physical properties of the cells, such as height, direction, and tilt. However, the Voronoi method and each of the aforementioned variations can also be used here.

The Bayesian framework can be extended iteratively with modules based on available auxiliary data. One such module is the Timing Advance module, which incorporates a variable from which the distance between a mobile device and its serving cell can be estimated. 

The outcome of the Bayesian framework is the \defn{location posterior}, which is an estimate of the location of a device given that it is connected to a certain cell. This location is not a single point, but a geospatial distribution. For implementation, we recommend to overlay the area of interest with a grid of square tiles, typically 100 by 100 meters. The location posterior then defines the probability that a device is located in a grid tile given that it is connected to a certain cell. This location posterior can be used for further statistical inference.


A similar Bayesian location estimation approach has been developed simultaneously with our work~\cite{ogulenko20}. Though their overall approach seems similar at first glance, there are a two significant differences. First, the proposed connection likelihood relies on aggregated Timing Advance data, which are often unavailable. The connection likelihood method that we propose uses cell plan data which are always available because these data are needed to build and maintain the mobile communication network infrastructure. When (aggregated) Timing Advance data are available, they can be used in our Timing Advance module. Second, the authors of~\cite{ogulenko20} advise against using a priori information such as land use, raising the argument that it is not objective. We argue that there is no such thing as an objective prior and will show that a non-uniform prior has added value for estimating the geographic location of devices.


The outline of the paper is as follows. In Section~\ref{section:modularframework} we describe our modular Bayesian framework. The location prior, connection likelihood, and timing advance module will be discussed in Sections~\ref{section:prior}, ~\ref{section:likelihood}, and~\ref{section:ta} respectively. A small example of the framework will be provided in Section~\ref{section:example}. How this framework can be used in statistical inference will be discussed in Section~\ref{section:statinf}. We introduce the signal strength model that can be used for the connection likelihood in Section~\ref{section:sigstr}. In Section~\ref{section:application} we illustrate our approach with an fictional application. The implementation is described in Section~\ref{section:implementation}. We conclude with a couple of remarks in Section~\ref{section:concludingremarks}.

\section{A modular Bayesian framework}\label{section:modularframework}
In the proposed method, we will overlay the geographic area of interest with a grid. The main advantage of using grid tiles is that different geospatial datasets can be combined without the need to calculate spatial intersections, which is a time consuming operation. Moreover, the mathematics described below is easier since all grid tiles have the same area.

The key to the proposed localisation method is Bayes' formula, which is used in the following way:
\begin{equation}
\centering
\label{eq:p_g_a}
\Prob(g \mid a) \propto \Prob(g) \Prob(a \mid g),
\end{equation} 
where $g$ represents a grid tile and $a$ a cell. The probability $\Prob(g)$ that a device is located in grid tile $g$ without any connection knowledge represents the location prior about the relative frequency of events at grid tile $g$. The connection likelihood $\Prob(a \mid g)$ is the probability that a device is connected to cell $a$ given that the device is located in grid tile $g$. The location posterior $\Prob(g \mid a)$ represents the probability that a device is located in grid tile $g$ given that the device is connected to cell $a$. 

As a technical aside, we note that the expression $\Prob(a \mid g)$ plays two distinct roles in our discussion. When we model $\Prob(g \mid a)$ via Eq.~\eqref{eq:p_g_a}, the cell $a$ is considered fixed. Hence $\Prob(a \mid g)$ is not a probability distribution, but a likelihood function with parameter $g$. On the other hand, when we will propose a model for $\Prob(a \mid g)$ in Section~\ref{section:likelihood}, it will be considered as a probability distribution over $a$ for fixed $g$. Although this distinction is not critical for the reader's understanding of our model, the intended interpretation should be clear from context. For simplicity, we will refer to $\Prob(a \mid g)$ as the connection likelihood henceforth.

\begin{figure}
\centering
\includegraphics[scale = 0.8]{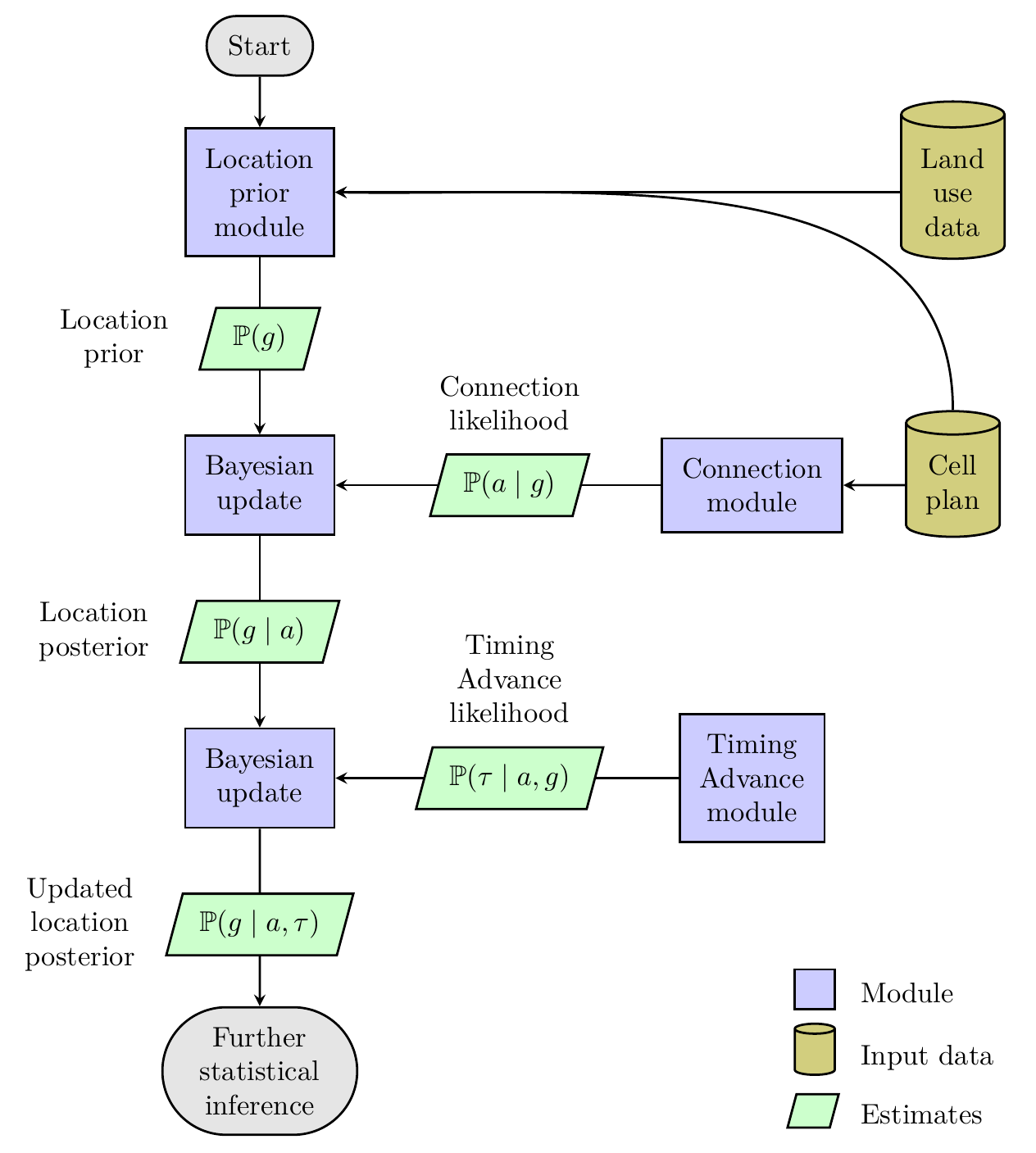}
\caption{A modular framework for modelling location posteriors.}
\label{fig:flowchart-modularsystem}
\end{figure}

The position of Eq.~\eqref{eq:p_g_a} in our modular Bayesian framework is illustrated in Fig.~\ref{fig:flowchart-modularsystem}. The location prior and connection likelihood are estimates produced from models we call the location prior module and connection module, respectively. The former module may use land use data as input, while we propose to use cell plan data. The location prior is then updated by the connection likelihood to form the location posterior. The localisation method can at this point be considered as complete and this posterior may be used for further statistical inference. Alternatively, the posterior can be seen as a new prior and updated again with other likelihoods if the necessary information is available. Fig.~\ref{fig:flowchart-modularsystem} shows how a Timing Advance module might be used to construct an updated location posterior.

The rest of this section is devoted to proposals for, and elaborations of these modules.


\subsection{Location prior}\label{section:prior}
We define the \defn{location prior} $\Prob(g)$ as the probability that a device is present in grid tile $g$, such that
\begin{equation}
    \sum_{g \in \mathcal{G}} \Prob(g) = 1,
\end{equation}
where $\mathcal{G}$ is the set of all possible location estimates in our model, in other words, the whole grid.

The definition of the location prior function will be based on assumptions about where a device is expected to be. In this section, we propose four options: the uniform prior, the land use prior, the network prior, and the composite priors. 



\subsubsection{Uniform prior}
When we use the \defn{uniform prior}, we assume the probability of a device being in any grid tile is the same value for every grid tile:
\begin{equation}
    \Prob_\uniform(g) := \frac{1}{|\mathcal{G}|}.
\end{equation}
A uniform prior is sometimes viewed as uninformative. In the case of mobile phone data, however, the implicit assumption that any grid tile is as likely as the next can lead to an underestimation of devices in urban areas and an overestimation of devices in rural areas. We therefore advise against using the uniform prior as a default prior without consciously assessing the plausibility of the underlying assumption.


\subsubsection{Land use prior}
An alternative to the assumption of uniformity is to use administrative sources on land use for the location prior. One would for example expect more devices in an urban area than in a meadow. Hence, our second proposed prior is called the \defn{land use prior} and it is based on a proportional expectation of the number of devices $n(g)$ in grid tile $g$ such that:
\begin{equation}
    n(g) \propto \mathbb{E}[\text{number of devices in $g$}] \quad \text{for all } g \in \mathcal{G}.
\end{equation}
The land use prior is then defined as:
\begin{equation}
    \Prob_\landuse(g) := \frac{n(g)}{
    \sum_{g^\prime \in \mathcal{G}} n(g^\prime)}.
\end{equation}
Due to the normalisation in its definition, the land use prior does require an explicit estimate of the number of devices per grid tile. Any proportional measure has the same effect. One way to utilise this is when information is available on land use classes of the grid, such as levels of urbanisation. Let there be $K$ land use classes, each with their own relative expected number of devices: $u_1, u_2, \ldots, u_K$. Let $w_1(g), w_2(g), \ldots, w_K(g) \in [0,1]$ be the proportion of the grid tile that is covered by each class respectively, such that
\begin{equation}
    \sum_{\mathclap{k = 1, \ldots, K}} w_k(g) = 1 \quad \text{for all } g \in \mathcal{G}.
\end{equation}
Then $n(g)$ can be modelled as
\begin{equation}
    n(g) := \sum_{\mathclap{k = 1, \ldots, K}} u_k \cdot w_k(g).
\end{equation}
An example of a simple land use classification and the relative expected numbers is shown in Table~\ref{table:prior_land_use}.

\begin{table}
\caption{An example of land use classes and their relative expected number of devices.} 
\begin{minipage}{\linewidth}
\footnotesize
\begin{center}
\begin{tabular}{lr}
\toprule
Land use class & $u_k$ \\
\midrule
Urban & 1.0 \\
Main roads & 0.5 \\
Other land & 0.1 \\
Water & 0.0 \\
\bottomrule
\end{tabular}
\end{center}
\end{minipage}
\label{table:prior_land_use}
\end{table}

One of the downsides of the land use prior is that the assumptions based on administrative sources are less flexible in the case of major events. A festival in a location which would ordinarily be a quiet meadow, but suddenly contains many devices, is not accounted for in the land use prior. Such events can be recognized by the positioning and setup of the cells. For instance, extra small cells are often used to compensate for large numbers of devices~\cite{wang15, tolstrup15}.

It can also be worthwhile to let the land use prior depend on the time and day. For instance, the expected number of devices in industrial areas might be smaller during nighttime and weekends compared to daytime and working days. 

\subsubsection{Network prior}

The following prior, which we call the \defn{network prior}, is defined as 
\begin{equation} \label{eq:prior_network} 
\Prob(g) := \frac{\sum_{a \in \mathcal{A}} s(g,a)}{\sum_{a \in \mathcal{A}} \sum_{g^\prime \in \mathcal{G}} s(g^\prime, a)},
\end{equation}
where $s(g, a)$ for a grid tile $g$ and cell $a$ is called \defn{signal dominance}, and $\mathcal A$ denotes the set of all cells in the network. In Section~\ref{section:likelihood}, a general notion of \defn{signal dominance} will be introduced to serve as input for our model for the connection likelihood. A specific instance of it, the \defn{Voronoi signal dominance} $s_\Vor(g, a)$, will be defined in that section as well, and another instance $s_\strength(g, a)$ will be elaborated on in Section~\ref{section:sigstr}. Any of these instances can be used for Eq.~\eqref{eq:prior_network}.

This prior, together with our model for the connection likelihood $\Prob(a \mid g)$, which is discussed in Section~\ref{section:likelihood}, simplifies Eq.~\eqref{eq:p_g_a} to
\begin{equation}
\label{eq:post_prop_sigdom}
\Prob(g \mid a) \propto s(g, a).
\end{equation}
The interpretation of the prior~\eqref{eq:prior_network} depends on the instance of the signal dominance $s(g, a)$ one has chosen. Let $F\colon \mathcal G \to \mathcal A$ be an arbitrary function that maps grid tiles to cells. Then, by setting
\begin{equation}
s_F(g,a) :=
\begin{cases}
1 & \text{if $a = F(g)$,} \\
0 & \text{otherwise},
\end{cases}
\end{equation}
the prior in Eq.~\eqref{eq:prior_network} becomes uniform. Since $s_\Vor$ is a specific instance of the function $s_F$, the network prior with $s=s_\Vor$ also simplifies to the uniform prior.

When using the term \defn{network prior} from here on, we refer to Eq.~\eqref{eq:prior_network} with signal dominance based on the instance $s_\strength(g, a)$.


Basically, the network prior reflects the distribution of the total signal over all the grid tiles.
This prior contains implicit knowledge about where an MNO is expecting people. The placement of cells is not without reason; generally, more cells are placed in crowded areas, such as city centers, than in quiet rural areas. Note that we could have defined the network prior using the cell density. However, since the network capacity also depends on the type and configuration of the cells and on the environment (buildings and trees will generally have a negative effect on the propagation of the signal) we use the signal dominance, through which these aspects are taken into account.

There are two aspects to be aware of when using the network prior. First, the placement of cells is based on estimated peak traffic rather than the average expected number of devices. MNOs normally provide better network coverage in railway stations than in residential areas, since the estimated peak traffic is higher; people typically use their phone more actively in railway stations and moreover, the expected number of devices fluctuates more over time. The second aspect to be aware of is that MNOs might place extra (partially overlapping) cells in an area for reasons other than an expected increase of the number of devices. They might do this, for example, when they detect that the quality of the network connection is insufficient on certain sections of land. In summary, the total signal strength of the network does not always reflect the estimated number of devices.


From a Bayesian perspective, it may seem odd to use the same input, i.e. the signal strength, in both the connection likelihood distribution and the location prior. However, we use it in two different, complementary, ways. The example in Section ~\ref{section:example} includes calculations in which the signal strength model is used for both the connection likelihood and location prior.


\subsubsection{Composite priors}
Our fourth and final proposed prior is less theoretically substantiated and more driven by practical considerations. One can combine all three priors described earlier as follows:
\begin{equation}
\begin{aligned}
    \Prob_\composite(g)
        &:= \pi_\uniform \cdot \Prob_\uniform(g) + {} \\
        &\phantom{{}:={}} \pi_\landuse \cdot \Prob_\landuse(g) + {} \\
        &\phantom{{}:={}} \pi_\network \cdot \Prob_\network(g),
\end{aligned}
\end{equation}
where $\pi := (\pi_\uniform, \pi_\landuse, \pi_\network)$ is any vector in $\RR^3$ such that $0 \leq \pi_\uniform \leq 1$, $0 \leq \pi_\landuse \leq 1$, $0 \leq \pi_\network \leq 1$, and $\pi_\uniform + \pi_\landuse + \pi_\network = 1$.

The components of $\pi$ represent the contributions of the three previously defined priors to the final \defn{composite prior} $\Prob_\composite(g)$. Hence both the advantages and disadvantages of all three priors are mixed. The optimal choice of $\pi$ depends on the situation at hand and must be derived experimentally. A composite prior is harder to interpret from a theoretical point of view compared to the three priors discussed so far, which can be seen as an additional disadvantage.


\subsection{Connection likelihood}\label{section:likelihood}

We define the \defn{connection likelihood} $\Prob(a \mid g)$ for a cell $a$ and a grid tile $g$ to be the probability that when a device located in grid tile $g$ generates an event at some cell, it does so at $a$. We model this probability as
\begin{equation}\label{eq:p_a_g}
    \Prob(a \mid g) := \frac{s(g,a)}{\sum_{a^\prime \in \mathcal{A}} s(g,a^\prime)},
\end{equation}
where $\mathcal{A}$ is the set of all cells in the MNOs network and $s(g,a) \in [0, \infty)$ stands for the \defn{signal dominance} (an umbrella term introduced by ourselves) received in grid tile $g$ from cell $a$. That is, the connection likelihood is the ratio of the signal dominance received from cell $a$ to the total value of signal dominance received from all cells. Different choices for modelling the signal dominance are possible, and any choice defines the \defn{connection module} in Fig.~\ref{fig:flowchart-modularsystem}. Note that $\Prob(a \mid g)$ is independent of rescaling the function $s(g, a)$ by a constant, and our convention is that $s(g, a)$ should be defined so as to take on values in the interval $[0, 1]$.

One simple method to define the connection module is via Voronoi tessellation. In this case, $s(g, a)$ is set to be
\begin{equation}\label{eq:s_vor}
    s_\Vor(g,a) := 
    \begin{cases}
    1&\text{if } g \in \Vor(a), \\
    0&\text{otherwise,}
    \end{cases}
\end{equation}
where $\Vor(a)$ is the set of grid tiles of which the centroids lie in the Voronoi region surrounding cell $a$. Denoting~\eqref{eq:p_a_g} in this case by $\Prob_\Vor(a \mid g)$ then gives 
\begin{equation}\label{eq:pag_vor}
\Prob_\Vor(a \mid g) = s_\Vor(g, a),
\end{equation}
and combining this connection likelihood with the uniform prior $\Prob_\uniform(g)$ results in the location posterior
\begin{equation}
\Prob_\Vor(g \mid a) =
    \begin{cases}
    \bigl\lvert\{g' \in \mathcal G \mid g' \in \Vor(a)\}\bigr\rvert^{-1} & \text{if $g \in \Vor(a)$,} \\
    0 & \text{otherwise.}
    \end{cases}
\end{equation}
In Section~\ref{section:sigstr} we propose a different, more advanced definition of the connection module by first approximating the signal strength $S(g,a)$ measured in $\si{\dBm}$, and then applying a transformation to it to obtain a signal dominance $s_\strength(g,a)$.


\subsection{Incorporating Timing Advance data}\label{section:ta}

Some MNOs include in their signalling data a so called \defn{Timing Advance} variable~\cite{3gpp-ts-36.213}. Timing advance is also known as round-trip time~\cite{zang10}, or trip-time bins~\cite{ogulenko20}. The value of such a variable represents a time duration, and it is used to estimate and adjust for communication delays. However, in combination with the speed of radio waves it can alternatively be used to estimate the distance between device and cell~\cite{kreher2011}. In the case of 4G signalling data the Timing Advance variable takes on values in the discrete set $\{0, 1, \ldots, 1282\}$. If an event then contains such a value $\tau$ for this variable, the associated device is located approximately in an annulus centered around the antenna of width \SI{78}{\metre} and an inner circle of radius $\tau \cdot \SI{78}{\metre}$.

One should be aware, though, of errors present in this approximation. As the authors of~\cite{raitoharju2010} observed, the Timing Advance variable may take on different values while the device has the same distance to the cell, and even when the device is at the same location at different times.

Knowledge of $\tau$ may be used to improve the location posterior $\Prob(g \mid a)$ through Bayesian updating:
\begin{equation}\label{eq:p_g_a_tau}
\Prob(g \mid a, \tau) \propto \Prob(g \mid a) \Prob(\tau \mid a, g),
\end{equation}
where the \defn{Timing Advance likelihood} $\Prob(\tau \mid a, g)$ can be modelled as the fraction of the grid tile $g$ which lies in the annulus around the cell $a$ specified by $\tau$. Since the maximum value of $\tau$ is $1282$, this new location posterior $\Prob(g \mid a, \tau)$ equals $0$ for grid tiles further than approximately \SI{100}{\kilo\metre} from the cell.

Computing the likelihoods $\Prob(\tau \mid a, g)$ for all Timing Advance annuli around all cells $a$ in the network and all tiles $g$ in the grid that is used might prove to be too expensive computationally if calculated in the way we suggest above. One could therefore model $\Prob(\tau \mid a, g)$ more coarsely as being $1$ if the centroid of $g$ lies in the annulus specified by $\tau$, and $0$ otherwise. If the grid tiles used are substantially larger than \SI{78}{\metre}, though, such as the $100$ by $100$ meter tiles we propose, this coarser Timing Advance likelihood model could increase the location estimation error for devices located in the smaller annuli. These errors can be mitigated by merging adjacent annuli. For example, one could model 
\begin{equation}\label{eq:p_tau_a_g}
    \Prob(\tau \mid a, g) := 
    \begin{cases}
    1&\text{if } g \in \Ann(\tau, b), \\
    0&\text{otherwise,}
    \end{cases}
\end{equation}
where $\Ann(\tau, b)$ specifies the grid tiles whose centroids lie in the annuli specified by $\{\tau-b, \ldots, \tau+b\}$, where $b$ is a globally defined integer, independent of $\tau$, $a$ and $g$, that determines how many annuli are merged on both sides to the annulus corresponding to $\tau$. 

Non-binary distributions can be constructed from Timing Advance data~\cite{ogulenko20}. The proposed connection likelihood relies on location of antenna, propagation direction and a PRACH curve, which is the total number of connections per month, per cell, and per Timing Advance distance band. PRACH curves can be incorporated in our Timing Advance module if $\Prob(\tau \mid a, g)$ is modelled as the observed ratio between devices that connect to Timing Advance annulus $\tau$ and all devices connected to antenna $a$.


\subsection{Example} \label{section:example}  

To illustrate the computations involved in the model, we consider a small fictional island of 1 by 3 kilometers. All numerical values mentioned in this example can be found in Table~\ref{table:island}. The island can be divided into three grid tiles of equal size, $g_1$, $g_2$ and $g_3$. Note that for a more realistic example we would use much smaller grid tiles, but for simplicity each tile in this example measures 1 by 1 kilometer.

\begin{figure}
\centering
\includegraphics[width=0.3\textwidth]{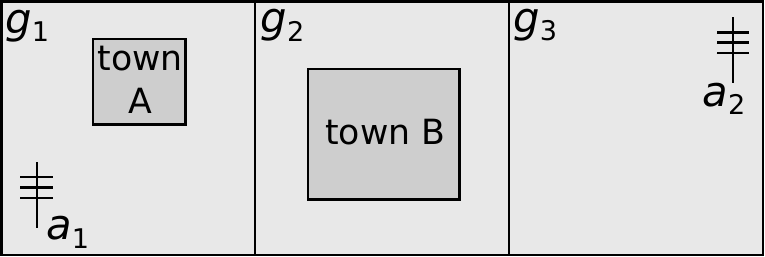}
\caption{A schematic top view of an island of 1 by 3 kilometers.}\label{fig:island}
\end{figure}

There are two towns A and B on the island, of which the latter is about three times as large as the former. Two cells, $a_1$ and $a_2$, have been installed. These are illustrated in Fig.~\ref{fig:island}. Cell $a_1$ has perfect signal dominance in $g_1$ and $g_2$, but no signal in $g_3$, while cell $a_2$ has perfect signal dominance in $g_3$ and $g_2$, but no signal in $g_1$. Perfect signal dominance and no signal are expressed by values $1$ and $0$, respectively. The signal dominance can be interpreted as the outcome of the signal strength model from Section \ref{section:sigstr}.

We calculate the four priors as defined in Section~\ref{section:prior} based on the above information. The connection likelihood for each cell is calculated according to Eq.~\ref{eq:p_a_g} based on the signal dominance values. Finally, the connection likelihood and priors are combined to location posteriors for all combinations of cells and priors. 

\begin{table}
\caption{The corresponding numbers of the fictional example where the composite prior is based on $\pi := (\pi_\uniform, \pi_\landuse, \pi_\network) = (0, \sfrac{1}{2}, \sfrac{1}{2})$.}
\begin{minipage}{\linewidth}
\footnotesize
\begin{center}
\begin{tabular}{llrrr}
\toprule
& & \multicolumn{3}{c}{Grid tile $g$} \\
\cmidrule{3-5}
& & $g_1$ & $g_2$ & $g_3$ \\
\midrule
Signal dominance        & $s(g, a_1)$ & 1 & 1 & 0\\
                        & $s(g, a_2)$ & 0 & 1 & 1\\
Location priors         &$\Prob_\uniform(g)$ & $\sfrac{1}{3}$ & $\sfrac{1}{3}$ & $\sfrac{1}{3}$\\
                        &$\Prob_\landuse(g)$ & $\sfrac{1}{4}$ & $\sfrac{3}{4}$ & 0\\
                        &$\Prob_\network(g)$ & $\sfrac{1}{4}$ & $\sfrac{2}{4}$ & $\sfrac{1}{4}$\\
                        &$\Prob_\composite(g)$ & $\sfrac{1}{4}$ & $\sfrac{5}{8}$ & $\sfrac{1}{8}$\\
Connection likelihood   &$\Prob(a_1 \mid g)$ & 1 & $\sfrac{1}{2}$ & 0 \\    
                        &$\Prob(a_2 \mid g)$ & 0 & $\sfrac{1}{2}$ & 1 \\
Location posterior      &$\Prob_\uniform(g \mid a_1)$ & $\sfrac{2}{3}$ & $\sfrac{1}{3}$ & 0\\
                        &$\Prob_\uniform(g \mid a_2)$ & 0 & $\sfrac{1}{3}$ & $\sfrac{2}{3}$\\
                        &$\Prob_\landuse(g \mid a_1)$ & $\sfrac{2}{5}$ & $\sfrac{3}{5}$ & 0\\
                        &$\Prob_\landuse(g \mid a_2)$ & 0 & 1 & 0\\
                        &$\Prob_\network(g \mid a_1)$ & $\sfrac{1}{2}$ & $\sfrac{1}{2}$ & 0\\
                        &$\Prob_\network(g \mid a_2)$ & 0 & $\sfrac{1}{2}$ & $\sfrac{1}{2}$\\
                        &$\Prob_\composite(g \mid a_1)$ & $\sfrac{4}{9}$ & $\sfrac{5}{9}$ & 0\\
                        &$\Prob_\composite(g \mid a_2)$ & 0 & $\sfrac{5}{7}$ & $\sfrac{2}{7}$\\
\bottomrule
\end{tabular}
\end{center}
\end{minipage}
\label{table:island}
\end{table}

The location posterior allows for all kinds of further calculations such as modelling the distribution of devices or even persons over grid tiles, when connection events are counted for each cell during a time interval. This process is complicated enough to be viewed as a separate research topic, especially when one is interested in probability distributions rather than point estimates. We consider it part of the \defn{further statistical inference} in the framework from Fig.~\ref{fig:flowchart-modularsystem}. 


\subsection{Statistical inference}\label{section:statinf}

The outcome of the modular system described in this paper is the location posterior $\Prob(g \mid a)$, which specifies the probability that a device is located in grid tile $g$, given that it is connected to cell $a$. This can be used to calculate the total number of devices that are present at a specific location during a specific time interval, or the number of devices that move from one city to another. However, many applications in official statistics are about numbers of people, for instance the number of visitors of a tourist destination during holidays, or the number of people who commute between two cities. Additional methods and auxiliary data are needed to translate estimates of devices to estimates of people.

A generic framework has been proposed to organize the production process needed for statistical inference on MNO data~\cite{ricciato20,salgado2020b}. According to this framework, the  production process runs through three distinct layers. The bottom layer is called the data- or D-layer and consists of the processing of raw mobile network data, which takes place at the MNO. The processing methods that take place in this layer are dependent on the mobile network technology used. The statistics- or S-layer is the top layer in which the processed mobile phone data is used for statistical purposes. The convergence- or C-layer connects these two layers with processing mobile network data sources into data that can be used for statistical purposes. This intermediate layer is needed since mobile network technology is complex and constantly changing. The output of the C-layer should be a stable source for the S-layer, in which this is used in combination with other data sources to produce statistics.

Our framework takes place in the D-layer, since mobile network data is processed for constructing the connection likelihood. Note that it does not matter which method is used for this process, since all described methods use mobile network data, e.g.\ the Voronoi method uses cell tower locations. The output of our framework, i.e.\ the location posterior, belongs in the C-layer, since this does not depend on technology, and hence can be used directly for statistical purposes. Using prior information could be theoretically be placed in the S-layer. Note that the process should ideally be run at the MNO due to potential privacy issues.

\section{Signal strength model}\label{section:sigstr}

This section describes the propagation of signal strength originating from a single cell. We distinguish two types of cells: omnidirectional and directional, resulting in two different propagation models. Omnidirectional cells have no aimed beam and their coverage area can be thought of as a circular disk. Directional cells point in a certain direction and their coverage area can be thought of as an oval with one axis of symmetry. In practice, small cells are usually omnidirectional and normal cells (i.e.\ attached to cell towers or placed on rooftops) are often directional~\cite{kora16}.

\subsection{Omnidirectional cells}
For omnidirectional cells, propagation of the signal strength $S(g,a)$ is modelled as
\begin{equation}
S(g,a) := S_0 - S_\dist(r_{g,a}),
\end{equation}
where $S_0$ is the signal strength at $r_0=1$ meter distance from the cell in $\si{\dBm}$ and $r_{g,a}$ is the distance between the middle point of grid tile $g$ and cell $a$ in meters. The value of $S_0$ can be different for every cell and is assumed to be a known property. In cell plan data, it is common to list the power $P$ of a cell in Watt, rather than the signal strength in $\si{\dBm}$. The value of $S_0$ can be calculated from $P$ using the conversion between Watt and $\si{\dBm}$~\cite{figueiras2010}:
\begin{equation}
S_0 = 30 + 10\log_{10}(P).
\end{equation}
The function $S_\dist(r)$ returns the loss of signal strength as a function of distance $r$:
\begin{equation} \label{eq:signal_radius}
S_\dist(r) := 10 \log_{10}(r^\gamma) = 10 \gamma \log_{10}(r),
\end{equation}
where $\gamma$ is the \textit{path loss exponent}, which resembles the reduction of propagation due to reflection, diffraction and scattering caused by objects such as buildings and trees~\cite{sri09}. In free space, $\gamma$ equals 2, but in practice higher values should be used. As a rule of thumb, 4 can be used for urban areas and 6 for indoor environments. Special situations, such as tunnels, could improve the propagation such that a value of less than 2 is applicable. The path loss exponent can be approximated by using the land use register.

\begin{figure}
\centering
\includegraphics[width=0.7\textwidth]{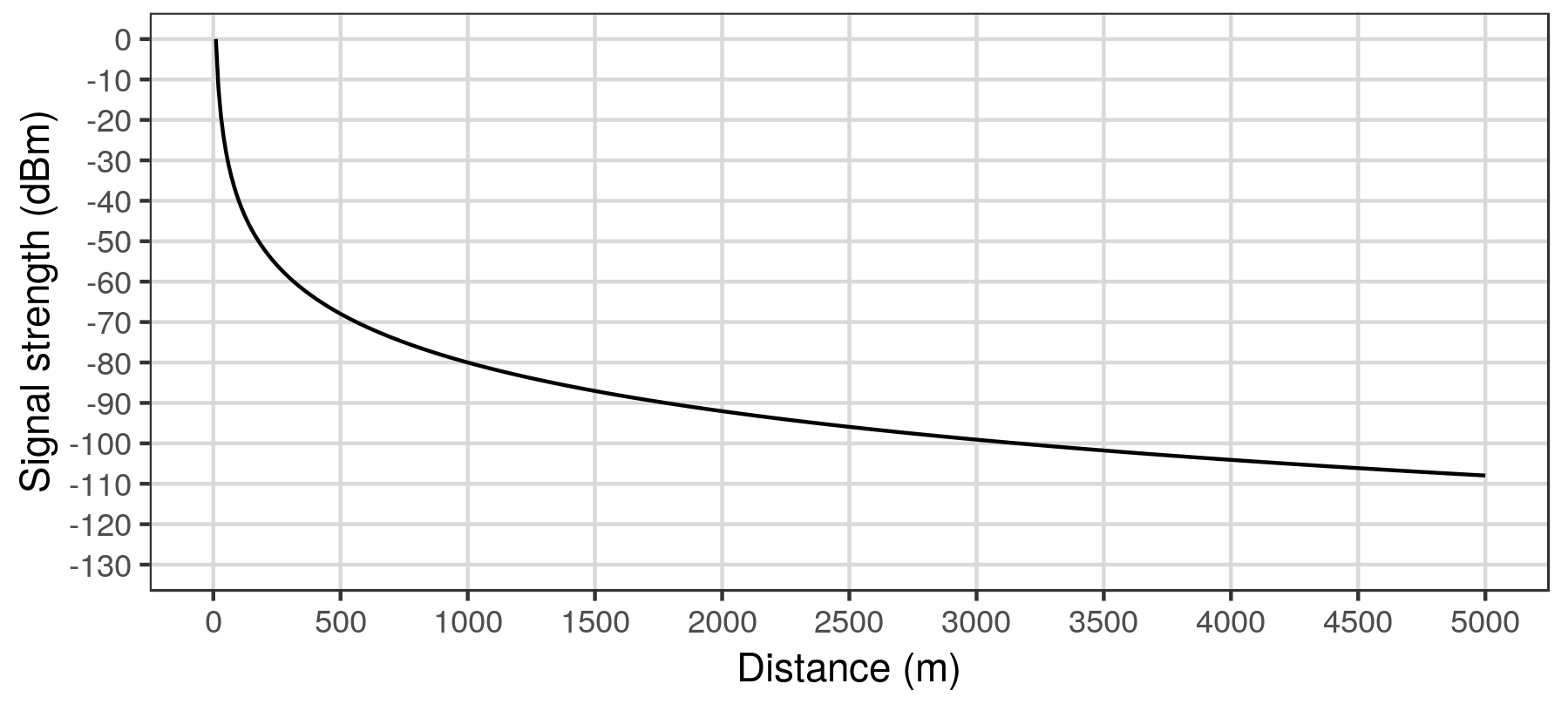}
\caption{Signal loss as a function of the distance for a specific cell.}\label{fig:dist}
\end{figure}

In Fig.~\ref{fig:dist}, the signal loss as a function of the distance is shown for a cell with \SI{10}{\watt} power that is standing in an urban environment ($\gamma=4)$.

\subsection{Directional cells}
A directional cell is a cell that is aimed at a specific angle. Along this angle, the signal strength is received at its best. However, the signal can also be strong in other directions. It is comparable to a speaker producing sound in a specific direction. The sound is audible in many directions, but is much weaker at the sides and the back of the speaker. We specify the beam of a directional cell $a$ by four parameters:

\begin{itemize}
\item The azimuth angle $\varphi_a$ is the angle from the top view between the north and the direction in which the cell is pointed, such that $\varphi_a \in [0, 360)$ degrees. Note that cell towers and rooftop cells often contain three cells with 120 degrees in between.

\item The elevation angle $\theta_a$ is the angle between the horizon plane and the tilt of the cell. Note that this angle is often very small, typically only four degrees. The plane that is tilt along this angle is called the \textit{elevation plane}.

\item The horizontal beam width $\alpha_a$ specifies in which angular difference from the azimuth angle in the elevation plane the signal loss is \SI{3}{\dB} or less. At \SI{3}{\dB}, the power of the signal is halved. The angles in the elevation plane for which the signal loss is \SI{3}{\dB} correspond to $\varphi_a \pm \alpha_a / 2$. In practice, these angles are around 65 degrees. 

\item The vertical beam width $\beta_a$ specifies the angular difference from $\theta_a$ in the vertical plane orthogonal to $\varphi_a$ in which the signal loss is \SI{3}{\dB}. The angles in which the signal loss is \SI{3}{\dB} correspond to $\theta_a \pm \beta_a / 2$. In practice, these angles are around 9 degrees.
\end{itemize}

Let $\delta_{g,a}$ be the angle in the elevation plane between the azimuth angle $\varphi_a$ and the orthogonal projection on the elevation plane of the line between the center of cell $a$ and the center of grid tile $g$. Similarly, let $\varepsilon_{g,a}$ be the angle from the side view between the line along the elevation angle $\theta_a$ and the line between the center of cell $a$ and the center of grid tile $g$. Note that $\varepsilon_{g,a}$ depends on the cell property of the installation height above ground level. We model the signal strength for directional cells as
\begin{equation}
S(g,a) := S_0 - S_\dist(r_{g,a}) - S_\azi(\delta_{g,a}, \alpha_a) - S_\elev(\varepsilon_{g,a}, \beta_a), 
\end{equation}
where $S_0$ is the signal strength at $r_0=1$ meter distance from the cell, in the direction of the beam so that $\delta=0$ and $\varepsilon=0$. The signal loss due to distance to the cell, azimuth angle difference and elevation angle difference is specified by $S_\dist$, $S_\azi$ and $S_\elev$, respectively. The definition of $S_\dist$ is similar to the omnidirectional cell and can be found in Eq.~\eqref{eq:signal_radius}. 



Each cell type has its own signal strength pattern for both the azimuth and elevation angles. These patterns define the relation between signal loss and the offset angles, i.e., $\delta_{g,a}$ for the azimuth and $\varepsilon_{g,a}$ for the elevation angles. We model the radiation pattern for both $S_\azi$ and $S_{elev}$ by a linear transformation of the Gaussian formula, each with different values for parameters $c$ and $\sigma$. Let
\begin{equation} 
f(\varphi) = c - c e^{-\frac{\varphi^2}{2 \sigma^2}},
\end{equation} 
where $c$ and $\sigma^2$ are constants, whose value is determined by numerically solving equations for a set of constraints. These constraints are different for $S_\azi$ and $S_\elev$ and depend on cell properties. 

\begin{figure}
\centering
\includegraphics[width=0.8\textwidth]{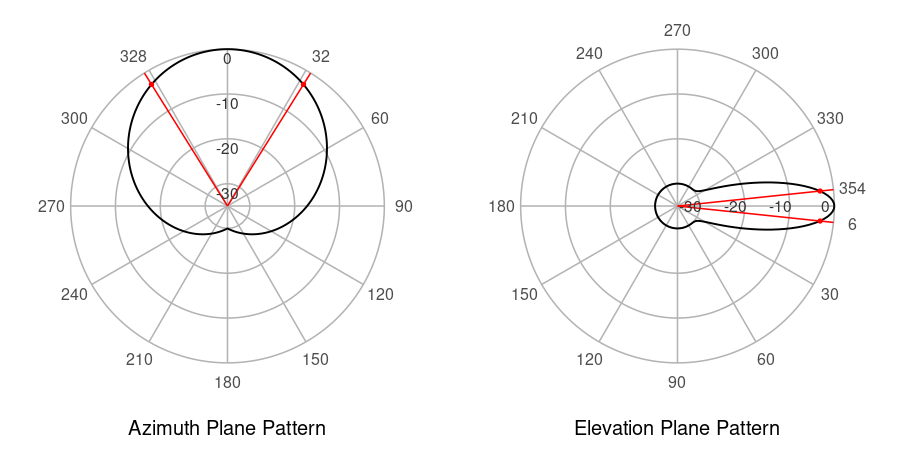}
\caption{Radiation patterns for the azimuth and elevation planes.}\label{fig:radiation}
\end{figure}

The resulting patterns are shown in Fig.~\ref{fig:radiation}. The black line shows the relation between signal loss and angle in the azimuth plane (left) and elevation plane (right). The grey circles correspond to the signal loss; the outer circle means \SI{0}{\dB} loss (which is only achieved in the main direction), the next circle corresponds to \SI{5}{\dB} loss, an so forth. The red lines denote the angles corresponding to \SI{3}{\dB} loss. The angle between the red lines is $2\alpha_a$ in the azimuth plane and $2\beta_a$ in the elevation plane.

Although these models approximate the general curve of real radiation patterns, the radiation patterns are more complex in reality, e.g.\ they often contain local spikes caused by so-called side and back lobes.

\begin{figure}
\centering
\includegraphics[width=1.0\textwidth]{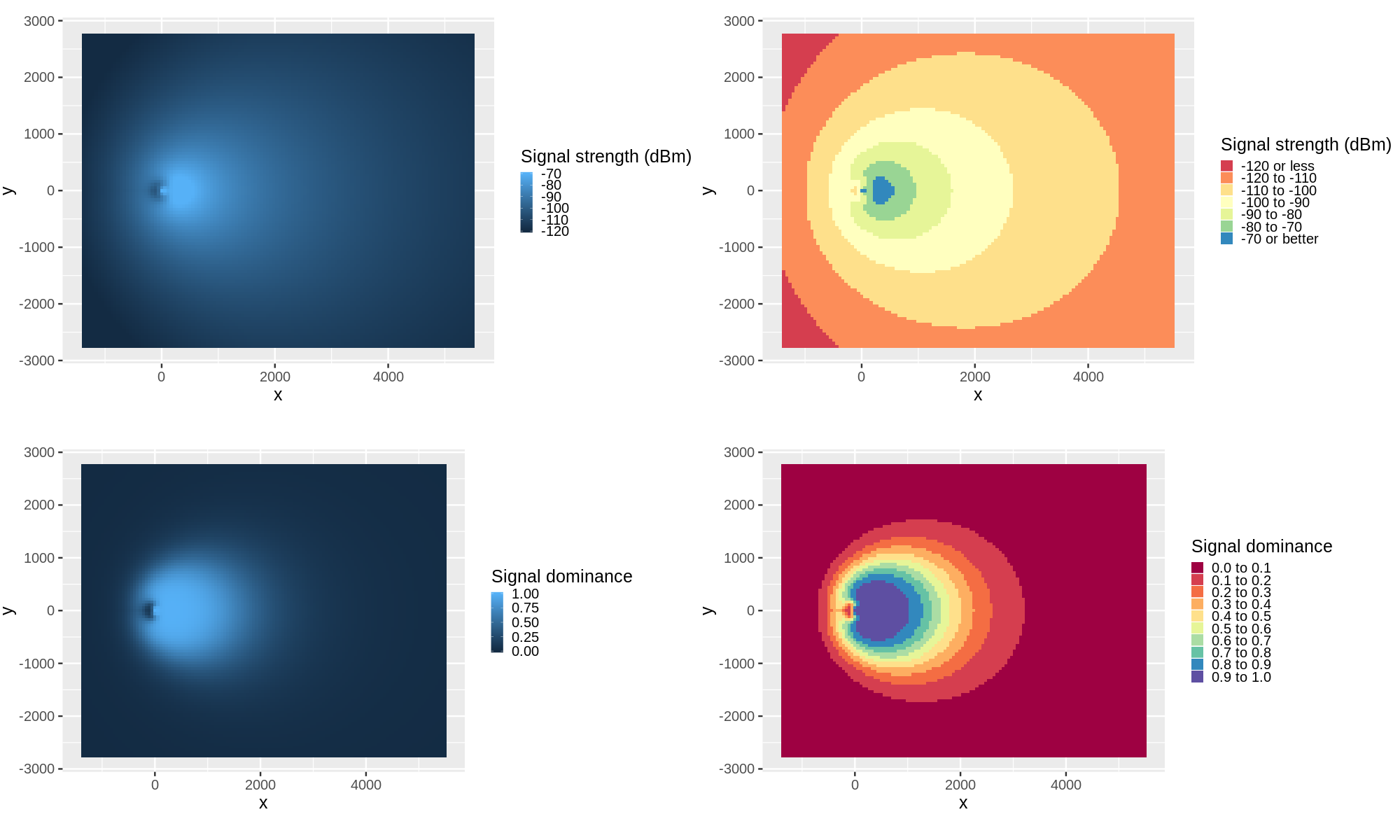}
\caption{Signal strength (top row) and signal dominance (bottom row) at ground level.}\label{fig:signal}
\end{figure}

Fig.~\ref{fig:signal} (top row) illustrates the signal strength at the ground level from above for a specific cell. In this case, the cell is placed at $x=0$, $y=0$ at 55 meters above ground level in an urban environment ($\gamma=4$), has a power of \SI{10}{\watt}, and is directed eastwards with an elevation angle (tilt) of 5 degrees, a horizontal beam width of 65 degrees and a vertical beam width of 9 degrees. Notice that the signal strength close to the cell, which on ground level translates to almost under the cell, is lower than at a couple of hundred meters distance. This is caused by relatively large $\varepsilon$ angles at grid tiles nearby the cell. 

\begin{table}
\caption{Indication of quality for signal strength in 4G networks.}
\begin{minipage}{\linewidth}
\footnotesize
\begin{center}
\begin{tabular}{ll}
\toprule
Signal strength (\si{\dBm}) & Quality \\
\midrule
$-70$ or higher & Excellent\\
$-90$ to $-70$ & Good\\
$-100$ to $-90$ & Fair\\
$-110$ to $-100$ & Poor\\
$-110$ or less & Bad or no signal\\
\bottomrule
\end{tabular}
\end{center}
\end{minipage}
\label{table:strength}
\end{table}

\subsection{Signal dominance}

The assignment of a cell to a mobile device does not only depend on received signal strength, but also on the capacity of the cells. The process of assigning devices to cells while taking into account the capacity of the cells is also called \defn{load balancing}.

Our model allows for two phenomena that we feel should not be overlooked. The first is the switching of a device when it is receiving a bad signal to a cell with a better signal. Table~\ref{table:strength} describes how the signal strength can be interpreted in terms of quality for 4G networks~\cite{kora16}. The second phenomenon is the switching between cells that is influenced by some decision making system in the network that tries to optimize the load balancing within the network. The specifics of this system are considered unknown. 

We assume that a better signal leads to a higher chance of connection. When a device has multiple cells available with a signal strength above a certain threshold, say \SI{-90}{\dBm}, the signal strengths are both more than good enough and the cell with the highest capacity is selected rather than the cell with the best signal strength. When the choice is between cells with a lower signal strength, one can imagine that their relative differences play a more important role in the connection process. 
However, when there are multiple cells available with a poor signal strength, it can be assumed that the signal strength value is less important than having capacity.
In short, we assume that signal strength plays a more important role in load balancing when it is in the middle range instead of in the high quality or low quality ranges.

\begin{figure}
\centering
\includegraphics[width=0.7\textwidth]{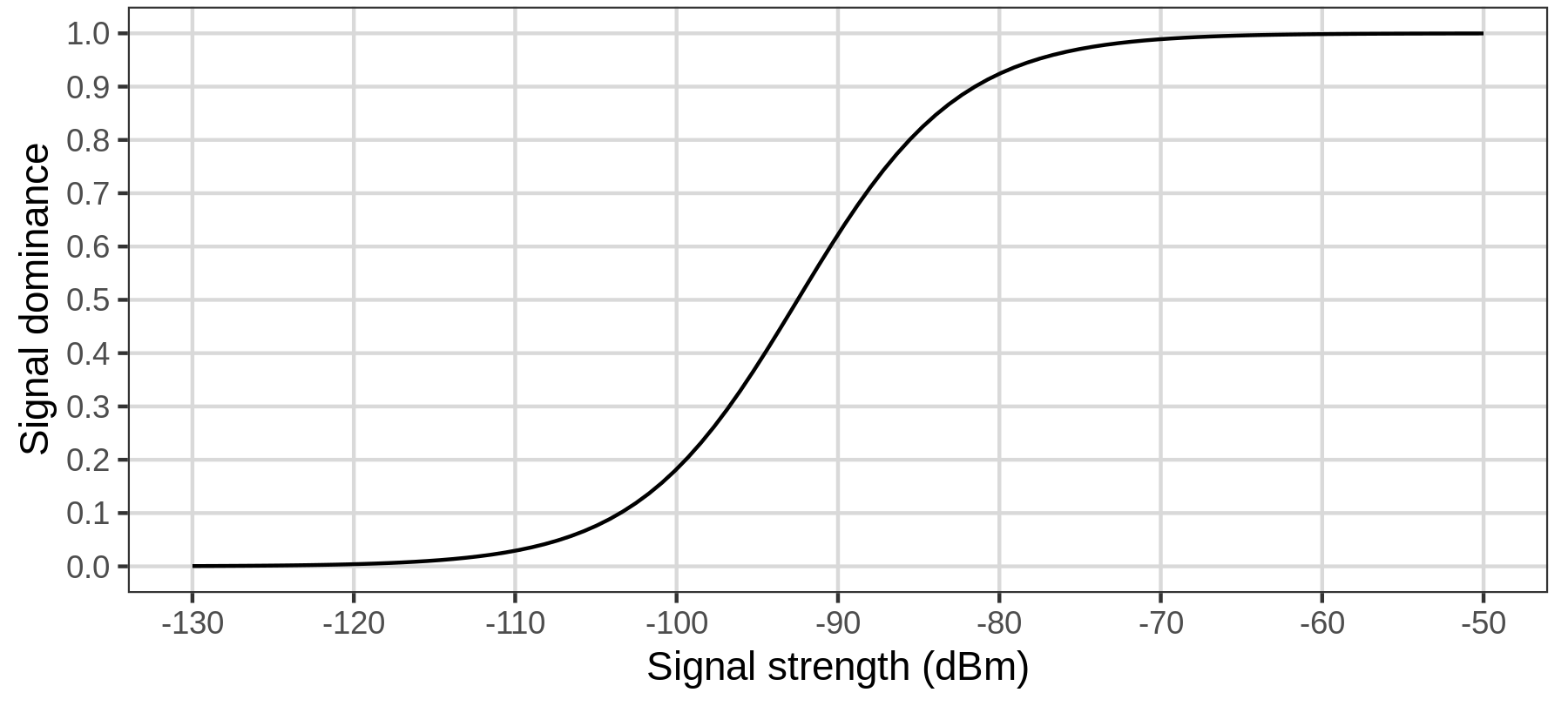}
\caption{Logistic relation between signal strength (\si{\dBm}) and signal dominance, where $S_\midp$ and $S_\steep$ are set to \SI{-92,5}{\dBm} and \SI{0.2}{\dBm} respectively to resemble Table \ref{table:strength}.}\label{fig:splot}
\end{figure}

To model this take on the load balancing mechanism, we use a logistic function to translate the signal strength $S(g,a)$ to the more interpretable signal dominance measure $s_\strength(g,a)$, which is then used to define the connection likelihood~\eqref{eq:p_a_g}. Let us define
\begin{equation}
\label{eq:rel_sign}
s_\strength(g,a) := \frac{1}{1 + \exp\bigl(- S_\steep \left(S(g,a) - S_\midp  \right)\bigr)},
\end{equation}
where $S_\midp$ and $S_\steep$ are parameters that define the midpoint and the steepness of the curve respectively. Fig.~\ref{fig:splot} shows an example of Eq.~\eqref{eq:rel_sign}.  

The signal dominance at ground level is shown in Fig.~\ref{fig:signal} (bottom row). The values that are shown are normalized by the sum of all values over all grid tiles, such that the normalized values form a probability distribution. Compared to the signal strength shown in Fig.~\ref{fig:signal} (top row), the signal dominance puts more emphasis on the geographic area that is in the range of the cell. Whether these signal dominance values resemble reality, should be validated by field tests.

\section{Application}\label{section:application}

\begin{figure}
\centering
\includegraphics[width=1\linewidth]{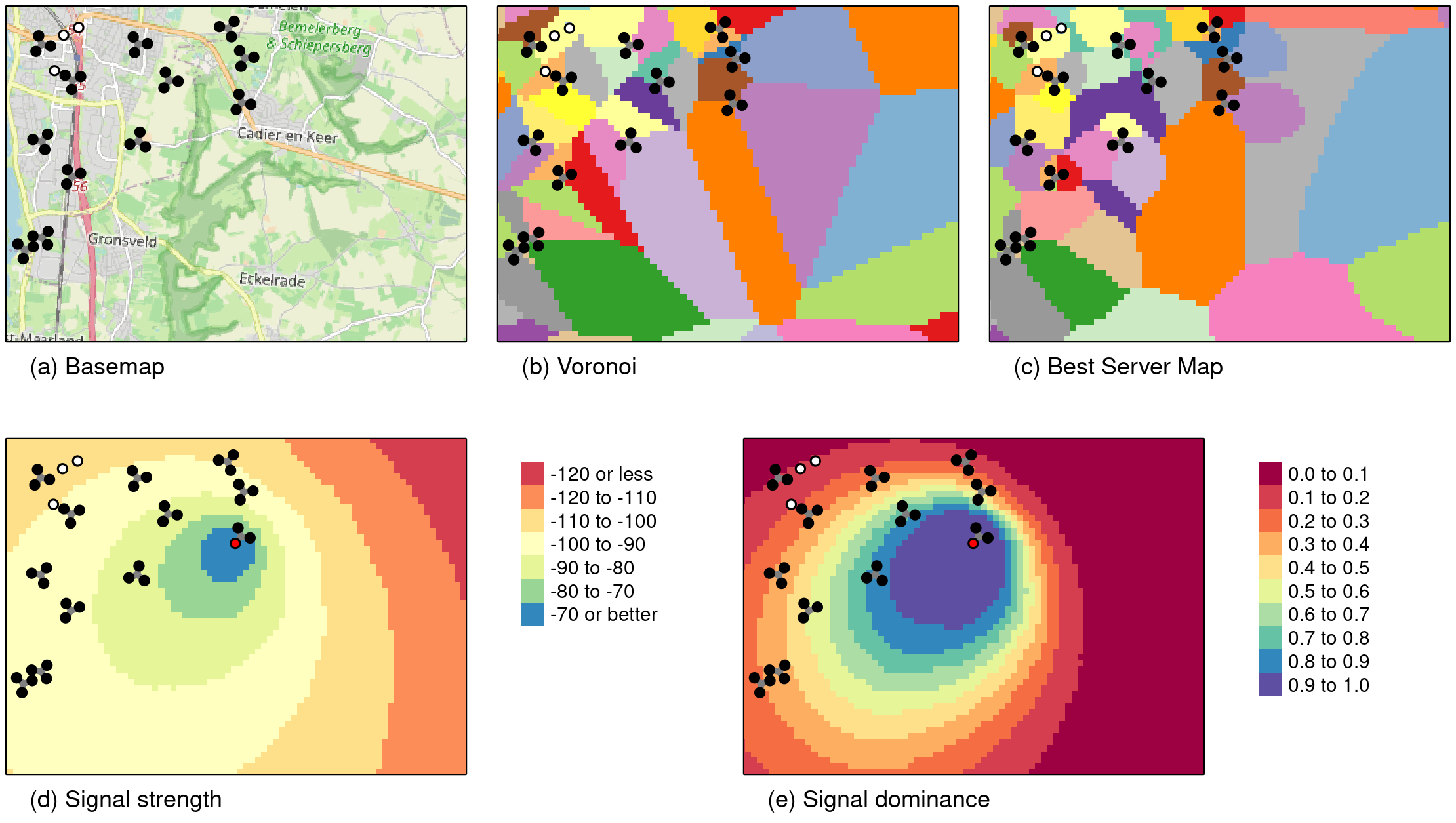}
\caption{The application of our signal strength model.} \label{fig:moblocbsm}
\end{figure}

A fictional example that resembles a real-world situation is illustrated in Fig.~\ref{fig:moblocbsm}(a), where a base map is shown with artificially placed cells. The northwestern part of this area is urbanised, while the rest is mostly rural with some small villages. The black dots represent directional macro cells, which are deployed in triplets, with propagation angles about 120 degrees apart from each other. The white dots represent omnidirectional small cells.

The Voronoi map for this example, which we will use as a reference for our signal strength model, is shown in Fig.~\ref{fig:moblocbsm}(b). We apply the Voronoi algorithm only to macro cells, where we shift the locations of the Voronoi points towards the propagation directions. Subsequently, the grid tile areas in which the small cells are located are subtracted from the Voronoi tessellation, and assigned to the small cells.

Fig.~\ref{fig:moblocbsm}(c) shows the \defn{best server map} based on our signal strength model. The best server map is a tessellation where each geographic location is assigned to the cell for which the signal strength is the strongest. 
This map provides insights in the results of our signal strength model. Observe for instance that some areas are relatively large and a few areas even consist of multiple disjoint patches. This is caused by a combination of factors, for instance the power, height, and elevation angle (tilt) of the cells, and the land elevation.

Figs.~\ref{fig:moblocbsm}(d) and (e) show the signal strength (in $\si{\dBm}$) and signal dominance ($s$) of one specific cell respectively, which is colored red. The propagation direction of this is south-west. Signal strength measurement tools can be used to validate the modelled signal strength, and if needed, to adjust the model parameters. The signal dominance is harder to validate, since it is a model of the deployed load balancing mechanism. A deep analysis of MNO data using ground truth location data, for instance GPS data, can be used for this validation task.

\begin{figure}
\centering
\includegraphics[width=1\linewidth]{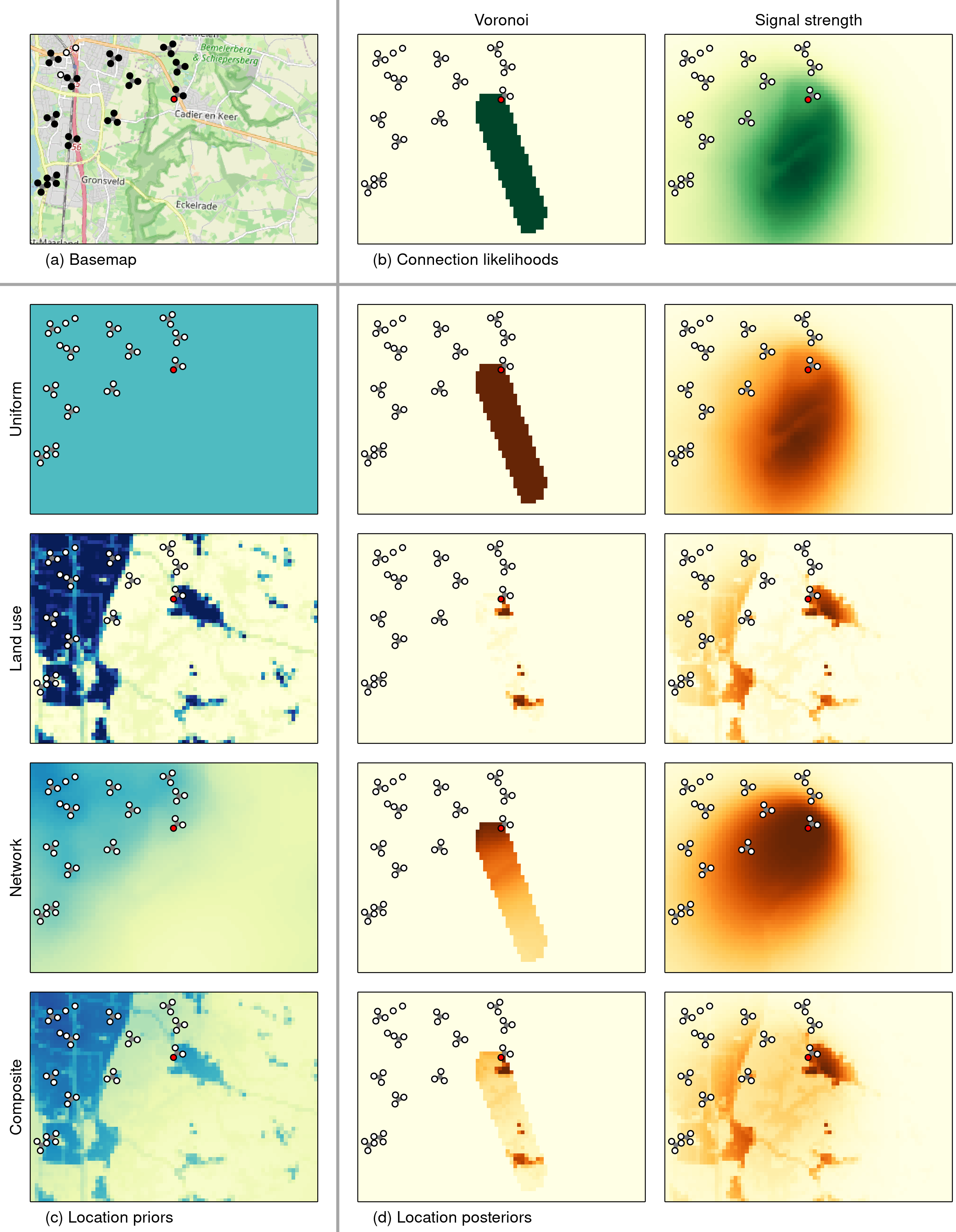}
\caption{Example showing how three location priors and two connection likelihoods are combined into six location posteriors.}\label{fig:moblocmultiple}
\end{figure}

Fig.~\ref{fig:moblocmultiple} shows the results of our modular framework for the same region, shown in (a). The rest of this figure is arranged as a cross table where the location priors are shown as rows in (c), the connection likelihoods as columns in (b), and the location posteriors as table cells in (d). For instance, the bottom right map of this figure is the result of using the composite prior (which consists of half the land use prior and half the network prior) and the signal strength likelihood.

In the maps shown in Fig.~\ref{fig:moblocmultiple}(b), (c), and (d), sequential color palettes have been applied where yellow corresponds to relatively low values and respectively dark green, blue, and brown correspond to relatively high values. For instance, the dark blue areas in the land use prior map correspond to the areas with buildings and roads,  and therefore, where devices are expected to be. The blue areas in the network prior map highlight the areas that have a high network coverage.

Two different connection likelihoods for the selected cell are shown in Fig.~\ref{fig:moblocmultiple}(b). Observe that the Voronoi likelihood has the same shape as the Voronoi area of this cell shown in Fig.~\ref{fig:moblocbsm}(b) in orange, which follows directly from Eq.~\eqref{eq:pag_vor} which states that a likelihood value equals 1 if and only if the corresponding grid tile is located in the Voronoi area. The high values for the signal strength likelihood are located in the south/south-west direction of the cell, whereas the propagation direction is south-west. This can be explained by the fact that there is less overlap with other cells in the south direction in comparison to the west direction.

The combination of the location priors and the connection likelihoods result in the location posteriors shown in Fig.~\ref{fig:moblocbsm}(d). The land use prior is seen to have a strong effect on the posterior distributions. The network prior places more weight on the areas with better network coverage. Note that, by applying Eq.~\eqref{eq:post_prop_sigdom}, the combination of network prior and signal strength likelihood results in a posterior distribution that is a rescaled version of the signal dominance, shown in Fig.~\ref{fig:moblocbsm}(e).
Finally, the combination of the composite prior and signal strength likelihood takes into account both land use and network coverage, and would therefore be a good candidate to be used for further statistical inference.

\clearpage

\begin{figure}
\centering
\includegraphics[width=1\linewidth]{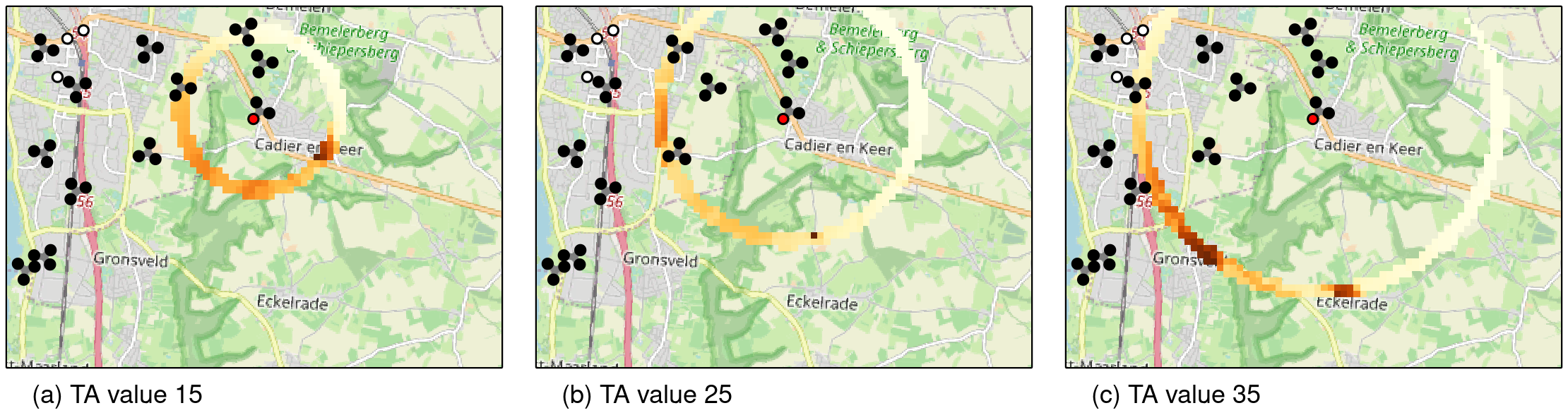}
\caption{Updated location posterior with Timing Advance.}\label{fig:moblocta}
\end{figure}

The final step in our modular framework is to update the location posterior by Timing Advance. Recall from Section~\ref{section:ta} that Timing Advance is a variable in MNO data from which the distance between the device and the cell can be estimated. Fig.~\ref{fig:moblocta} illustrates the location posterior shown in the bottom right map of  Fig.~\ref{fig:moblocmultiple} (composite prior, signal strength likelihood) updated with $\tau$ values of 15, 25, and 35. The Timing Advance likelihood defined by Eq.~\eqref{eq:p_tau_a_g} has been used with parameter $b$ set to 1, which means three annuli have been merged, in Fig.~\ref{fig:moblocta}(a) the annuli that correspond to $\tau$ values of 14, 15, and 16. In this application, the width of each annulus is 78.12 meter, so this means that only grid tiles whose centroids are between 1094 to 1328 away from the selected cell have a positive probability in the updated posterior. For Fig.~\ref{fig:moblocta}(b) and (c) those distance intervals are [1878, 2109] and [2656, 2890] respectively.

\clearpage

\section{Implementation}\label{section:implementation}

The methods described in this paper have been implemented in \texttt{mobloc}~\cite{mobloc} and \texttt{mobvis}~\cite{mobvis}, packages for the programming language \texttt{R}.
The former package is used for the calculations, the latter for the graphical user interfaces and visualizations. In this section we present a general work flow for \texttt{mobloc} and illustrate how \texttt{mobvis} is used in this work flow. Details and reproducible examples can be found in the documentation of the packages.

\begin{figure}
\centering
\includegraphics[width=0.8\textwidth]{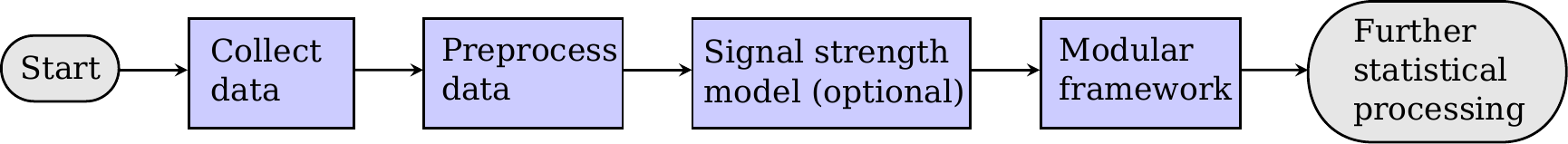}
\caption{Work flow of \texttt{mobloc}.}\label{fig:mobloc}
\end{figure}

The general work flow of \texttt{mobloc} is depicted in Fig.~\ref{fig:mobloc}. The first step in the work flow is to collect all relevant datasets. The most important dataset is is the cell plan. The geographical locations of the cells is required, whereas other physical properties, such as height, propagation direction, and power, are recommended in order to run the signal strength model. Other datasets that can be used as input are elevation data, land use data, and administrative region boundaries.

The collected data need to be in the correct format in order to be processed. Information about which object classes are supported are described in the package documentation~\cite{mobloc}. The preprocessing stage is also used to make sure that the spatial objects have one common Coordinate Reference System (CRS). It is strongly recommended to use a CRS that preserves areas and distances for the region of interest~\cite{maling92}. The function \texttt{validate\_cellplan} is used to validate the cell plan and to check whether it is consistent with the other input data objects.

\begin{figure}
\centering
\includegraphics[width=1.0\textwidth]{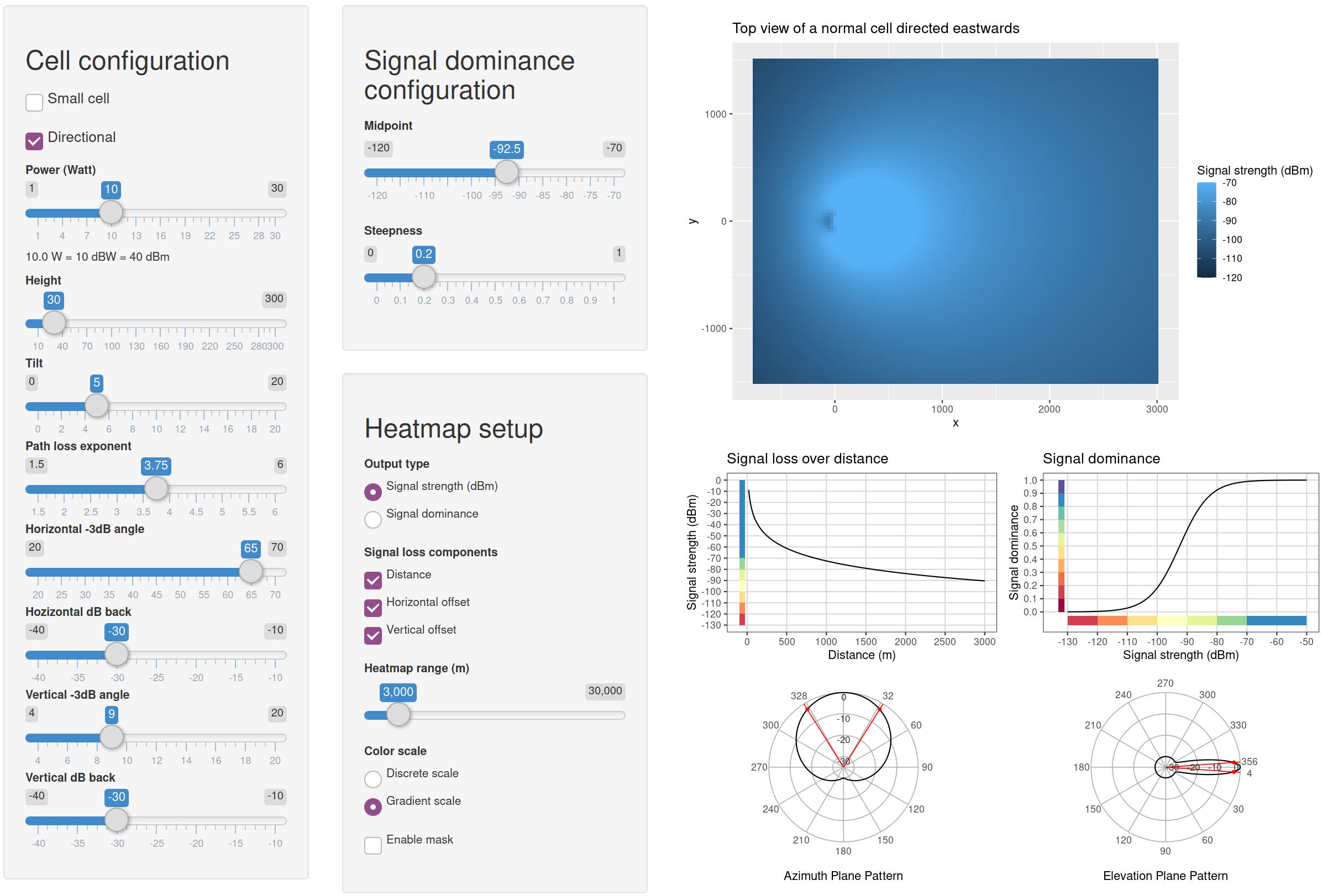}
\caption{Screenshot of the signal strength model setup tool.}\label{fig:prop}
\end{figure}

The third step in the work flow is to run the signal strength model. This step is only required if the network prior is used or if the signal strength model is used in the connection likelihood module. The parameters for the signal strength model can be configured with a graphical user interface started with the function \texttt{setup\_sig\_strength\_model} from the \texttt{mobvis} package. A screenshot of this tool is shown in Fig.~\ref{fig:prop}. The function \texttt{compute\_sig\_strength} from the \texttt{mobloc} package is used to compute the signal strength and the corresponding signal dominance.

\begin{table}
\caption{An overview of \texttt{mobloc} functions.}\label{table:functions}
\begin{minipage}{\linewidth}
\footnotesize
\begin{center}
\begin{tabular}{lll}
\toprule
Module & Option & Function \\
\midrule
Location prior & Uniform prior & \texttt{create\_uniform\_prior}\\
 & Network prior &  \texttt{create\_network\_prior}\\
 & Auxiliary data (e.g. land use) prior & \texttt{create\_prior}\\
Connection likelihood & Signal strength & \texttt{create\_strength\_llh}\\
& Voronoi & \texttt{create\_voronoi\_llh}\\
Bayesian update & Posterior $\Prob(g \mid a)$, see Eq.~\eqref{eq:p_g_a} & \texttt{calculate\_posterior}\\
 & Posterior $\Prob(g \mid a, \tau)$, updated & \texttt{update\_posterior\_TA}\\
 &  $\mbox{   }$ with Timing Advance, see Eq.~\eqref{eq:p_g_a_tau} & \\
\bottomrule
\end{tabular}
\end{center}
\end{minipage}
\end{table}

The fourth and final step is to run the whole modular framework that is depicted in Fig.~\ref{fig:flowchart-modularsystem}. The functions that correspond to the modules are listed in Table~\ref{table:functions}. 

\begin{figure}
\centering
\includegraphics[width=1.0\textwidth]{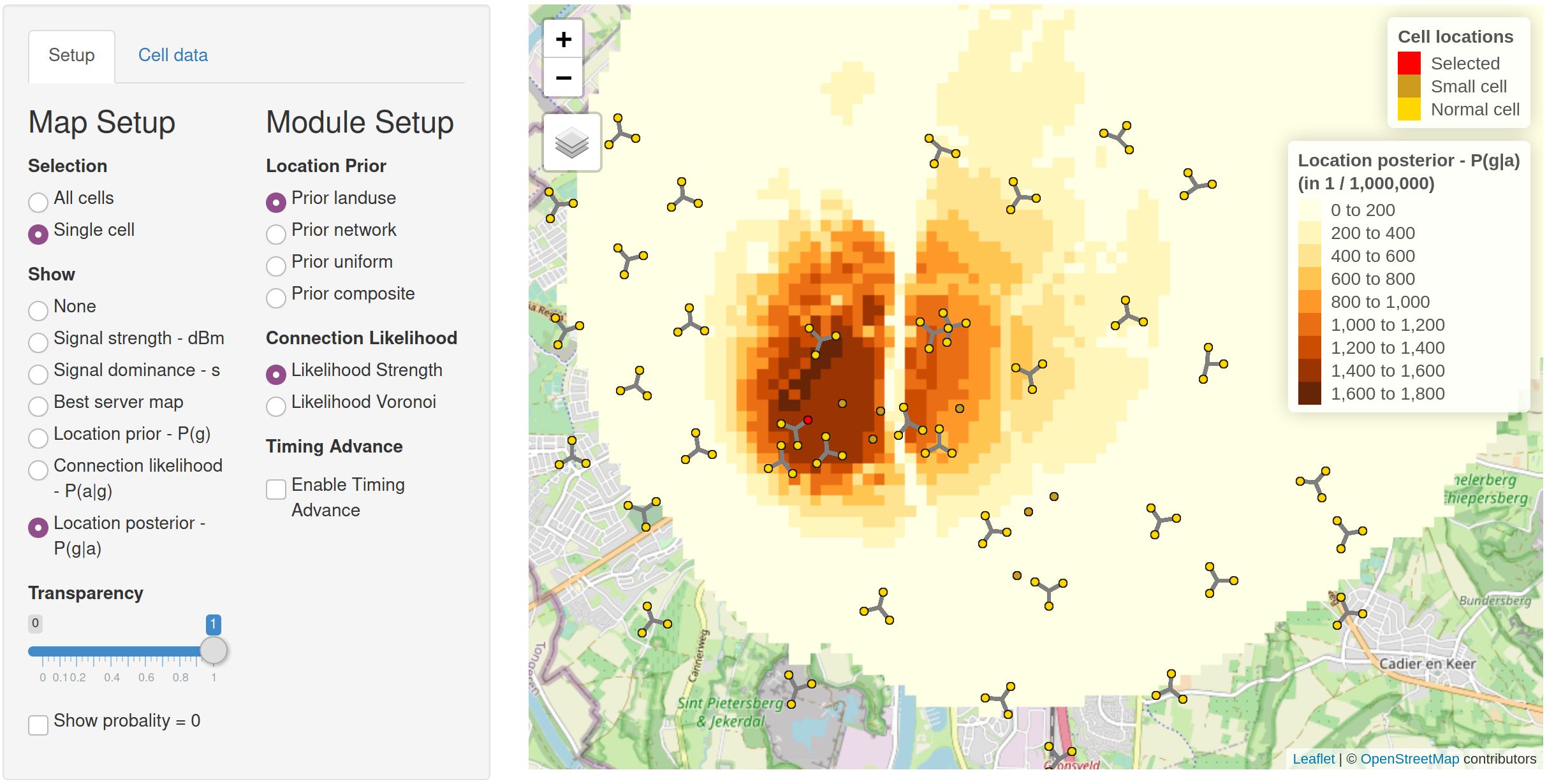}
\caption{Screenshot of the \texttt{mobloc} exploration tool.}\label{fig:explore}
\end{figure}

The \texttt{mobvis} package contain functions to visualize the results. All maps from Figs.~\ref{fig:moblocbsm},~\ref{fig:moblocmultiple}, and~\ref{fig:moblocta} have been created with \texttt{mobvis}. This package also contains an interactive tool, started with \texttt{explore\_mobloc}, in which the results are visualized. Via this tool, it is possible to select different cells, and compare different module options. A screenshot is shown in Fig.~\ref{fig:explore}, in which the same fictional cell plan data is used as described in Section~\ref{section:application}.


The output of \texttt{mobloc} is a data table with the following variables: cell identification number, tile identification number, optionally a Timing Advance value, and the corresponding posterior probability. This table can be used for further statistical processing and analysis with MNO data~\cite{ricciato20,salgado2020b}.

\section{Concluding remarks}\label{section:concludingremarks}
We have proposed a Bayesian approach to estimate the location of mobile devices using mobile operator network data. The methods described in this paper are modular, in the sense that one method can easily be replaced by another. If, for instance, a better propagation model exists, which for example uses a 3d model of the environment, this can be used together with the other methods. The same applies for the signal dominance function, connection likelihood, and location prior.


An MNO often facilitates mobile communication via multiple generations of networks (e.g. 3G and 4G). For each generation, an MNO maintains a cellular network. We currently assume that a mobile device only connects to cells from one network generation, that is, the latest generation which the device supports. When this assumption holds, the methods can be independently applied for each generation. The networks can be viewed as independent because a device will only connect to cells from one generation. In reality however, this assumption might not hold for reasons such as coverage gaps, capacity and network-specific optimal communication mode such as text, voice and internet. More research on switching between cells from different generations is needed.

For each generation, cells will serve at different frequencies. For instance an MNO may have a network of 4G cells that serve at 900, 1800 and \SI{2100}{\hertz}. As with handling of cells from multiple generations, more research is needed on the process of switching between frequencies.

Our propagation model can be calibrated using other data sources. Field measurements of received signal strength could provide insights in several parameters, for instance the power of the cells and the path loss exponent. The calibration process should ideally be executed for each mobile phone network, since they may be configured in different ways. 

Alternative estimation methods that build on our methods have been proposed recently. One method is the maximum likelihood estimator, which uses the connection likelihood~\cite{ricciato16,laan2019}. Our signal strength model can be used here for the connection likelihood, but this is not necessary. More recently, another method has been proposed that uses the connection likelihood to select a range of candidate estimates, and subsequently selects the estimate that best confirms prior information~\cite{ricciato20b}. 
Although the results of both methods are promising when applied to fictitious examples, they rely more on the connection likelihood than our method does. In scenarios where the connection likelihood is of lesser quality due to incomplete cell plan data or where prior data are available that have good predictive value, our location posterior estimation method may be preferable. Further research is needed to compare these methods by applying them with real data and validate them with external sources.

MNOs often have additional data sources that can be used for validation, such as coverage maps and best server maps, which are usually created with propagation models and measurements. One could compare these maps to those resulting from location estimation methods. However, be aware that these maps are often created for other purposes, such as network analysis and optimization. More research is needed on how to interpret the (dis)similarities between such maps.



A more thorough validation approach is to compare logged location data from a sample of devices, e.g. with GPS, and compare this to the location estimates that are based on MNO data from the same sample of devices. This would require consent both from the MNO as well as from the device owners.

The work of this paper has been focused solely on static situations; we estimate the location posterior probabilities of a device given the cell at which a certain event has been logged. However, when we consider the dynamic behaviour of a device and take into account the sequence of logged events, much more information can be extracted by interrelating the corresponding location posterior probabilities, for instance by using a Hidden Markov Model~\cite{salgado2020b}.

\bibliographystyle{josisacm}
\bibliography{main}

\end{document}